\documentclass[aps,prl,twocolumn,groupedaddress,floatfix]{revtex4-1}
\usepackage{amssymb,amsmath}
\usepackage{graphicx}
\usepackage{subfigure}
\usepackage[english]{babel}
\usepackage{float}
\usepackage{color}

\begin{document}

\title{All-optical logic gates in Stub photonic lattices}

\author{Basti\'an Real$^{1}$, Camilo Cantillano$^{1}$, Dany L\'opez-Gonz\'alez$^{1}$, Alexander Szameit$^{2}$, Masashi Aono$^{3}$, Makoto Naruse$^{4}$, Song-Ju Kim$^{5}$, Kai Wang$^{6}$, and Rodrigo A. Vicencio$^{1}$}

\affiliation{$^{1}$Departamento de F\'{\i}sica and MSI-Nucleus on Advanced Optics, Facultad de Ciencias, Universidad de Chile, Santiago, Chile}

\affiliation{$^{2}$Institute for Physics, University of Rostock, Albert-Einstein-Strasse 23, 18059 Rostock, Germany}

\affiliation{$^{3}$ Faculty of Environment and Information Studies, Keio University, 5322 Endo, Fujisawa, Kanagawa 252-0882, Japan}

\affiliation{$^{4}$ Network Research Institute, National Institute of Information and Communications Technology, 4-2-1 Nukui-kita, Koganei, Tokyo 184-8795, Japan}

\affiliation{$^{5}$ WPI Center for Materials Nanoarchitectonics, National Institute for Materials Science, 1-1 Namiki, Tsukuba, Ibaraki 305-0044, Japan}

\affiliation{$^{6}$Nonlinear Physics Centre, Research School of Physics and Engineering, The Australian National University, Canberra, ACT 2601, Australia}

\begin{abstract}

We experimentally study a Stub photonic lattice and excite their localized linear states originated from an isolated Flat Band at the center of the linear spectrum. By exciting these modes in different regions of the lattice, we observe that they do not diffract across the system and remain well trapped after propagating along the crystal. By using their wave nature, we are able to combine -- in phase and out of phase -- two neighbor states into a coherent superposition. These observations allow us to propose a novel setup for performing three different all-optical logical operations such as OR, AND, and XOR, positioning Flat Band systems as key setups to perform concrete applications at any level of power. 

\end{abstract}

\maketitle


\section{Introduction}

Very recently, a new topic has attracted a lot of attention from different physical communities working with lattices as a main framework for their studies. By using non standard geometries, different periodic lattices have been suggested to achieve the goal of localization without any linear or nonlinear impurity, and without any constraint on the level of power. Flat Band (FB) lattices possess a special linear spectrum where at least one of their linear bands is completely flat~\cite{luisFB}. This implies that the modes belonging to this special band do not diffract at all and remain localized in space as long as the system length. It has been shown~\cite{berg} that 2D FB linear modes consist on line extended states which could reduce abruptly its size by judicious linear combinations in different directions of the lattice. Therefore, it is possible to create a completely coherent linear base composed of very localized states. For example, a Lieb lattice~\cite{lieb1,lieb2,liebseba,liebchen} corresponds to a square depleted lattice, having three sites per unit cell, with two dispersive bands and one completely flat. Modes from this band have only four sites different to zero, being therefore perfectly localized in space. A Kagome lattice~\cite{kagima,kagchen} is also an interesting example, where the localized modes consist of only six sites different to zero. In all known FB systems, it is possible to find compact localized states, which occupy one or several unitary cells depending on the lattice geometry~\cite{flachFB}. These states are also exact solutions at the nonlinear regime, but they are not necessarily stable~\cite{kagnl,serbnl,sawnl,luisFB}.

Only few quasi one-dimensional FB systems have been experimentally explored. Two years ago, it was reported the observation of a localized state on a Diamond lattice, consisting on a localized mode presenting only two out of phase sites~\cite{olseba} (this is indeed the smaller FB state known in any system). Very recently, a Sawtooth lattice was tested experimentally~\cite{sawolus}, showing how the absence of transport is related to the appearance of a FB. Theoretical studies on Sawtooth lattices also explore quantum topological excitations~\cite{qte} and Bose--Einstein condensation~\cite{bec}. Polariton condensation was shown in 1D Lieb (Stub) lattices, in the context of micro-pillar optical cavities~\cite{prlamo}. This was the first experimental observation of properties related to Stub lattices, although the isolated excitation of a localized FB state was not possible on that experiment.

Some efforts have been focused on performing non-diffractive image transmission schemes using FB systems~\cite{kagima,kagchen,lieb2,liebchen}, although no logical operations have been suggested yet. Over the last two decades, various all-optical logic gates with and without semiconductor optical amplifiers (SOA) have been proposed, reflecting the demand to overcome the speed limit of electronic devices~\cite{AOT}. For example, all-optical NAND-NOT-AND-OR gates were proposed using optical fibers presenting gain and losses~\cite{daisyAO}. Interestingly, most of these previous studies tend focus on proposing quantum optical gates setups~\cite{qo1,qo2,qo3}, more than classical ones. Moreover, lattices have being poorly studied to implement logical operations in the linear regime, where most of the applications operate. Therefore, the ability of FB lattices to support linear localized states, which can be combined to form completely coherent -- time and space -- configurations, appears as an important problem to explore in more detail, in order to propose concrete all-optical solutions.

In this work, we study the fundamental properties of a Stub photonic lattice and its potential to developing novel all-optical logic gates without SOA. For the first time to our knowledge, we are able to excite a FB Stub linear localized mode. This state propagates without suffering diffraction and can be combined with neighbor modes to generate arbitrary linear combinations. We explore the implementation of three basic logical operations using the simple and powerful properties of a Stub FB photonic lattice. By superposing two FB modes we are able to generate the following logical gates: AND, OR and XOR. We use the wave nature of FB modes and combine them using different phases.

\section{Model}
\begin{figure}[t]
\begin{center}
\includegraphics[width=0.47\textwidth]{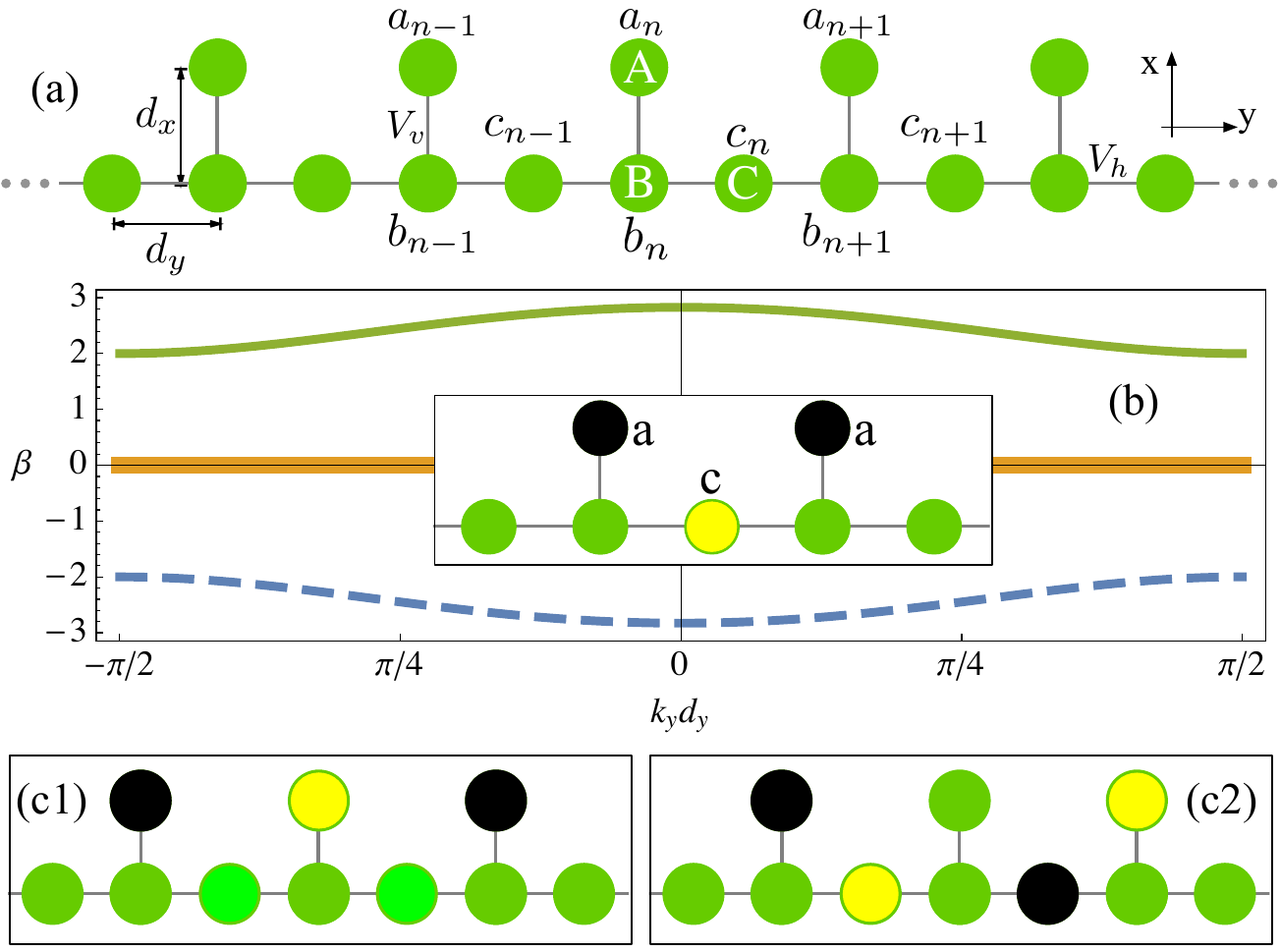}
\caption{(a) A Stub lattice configuration. (b) Linear bands for $V_h=1$, $V_v=2$. (b)-inset Flat Band compact state. Two FB states combined: (c1) in phase and (c2) out of phase.}\label{f1}
\end{center}
\end{figure}
%
A Stub photonic lattice consists of a main row with extra sites every two waveguides [see Fig.~\ref{f1}(a)]. Light evolution on this waveguide array occurs along the $z$ direction, which is orthogonal to the transversal periodic Stub structure. Based on coupled-mode theory, we describe the light dynamics by means of discrete linear Schr\"{o}dinger equations~\cite{rep1,rep2}, what in this lattice geometry reads as:
\begin{eqnarray}
-i\frac{d a_n}{d z} =V_v b_n\ ,\ \ \ -i\frac{d c_n}{d z} =V_h\left(b_{n+1}+b_{n}\right)\ ,\nonumber\\
-i\frac{d b_n}{d z} =V_v a_n+V_h\left(c_n+c_{n-1}\right)\ .
\label{model}
\end{eqnarray}
$a_n$, $b_n$ and $c_n$ represent the amplitude of the fundamental electric field mode at different waveguides forming the unitary cell [composed of sites $A$, $B$ and $C$ in Fig.~\ref{f1}(a)]. We assume only nearest neighbors interactions, governed by horizontal $V_h$ and vertical $V_v$ coupling constants, which are originated from a weak overlap between the fundamental modes of adjacent waveguides. In general, these coefficients depend on the separation between two waveguides [$d_x$ or $d_y$ in Fig.~\ref{f1}(a)], but also on the direction of coupling depending on a given waveguide profile~\cite{alex1}. We consider an homogenous array such that all the individual waveguides have the same propagation constants, therefore we eliminated this dependence in model (\ref{model}).

First of all, we solve model (\ref{model}) using the stationary ansatz: $\psi_n(z)=\psi_0 \exp{i(k_y n d_y-\beta z)}$, with $\psi_n(z)=a_n(z),\ b_n(z),\ c_n(z)$, depending on the particular waveguide position. This ansatz represents a plane wave travelling transversally in the array while propagating longitudinally along the $z$ direction. $k_y$ represents the horizontal wave vector. By inserting this ansatz, we get the linear bands (dispersion relation): $\beta=0,\ \pm \sqrt{4V_h^2\cos^2 (k_yd_y)+V_v^2}$. We plot the spectrum in Fig.~\ref{f1}(b) and observe two dispersive (and opposite in curvature) linear bands, and one completely flat at $\beta=0$. $b_n$ amplitudes are always zero for the linear modes belonging to this band. For a Stub lattice, a compact linear state is composed of only three sites different to zero~\cite{prlamo}. The relation between the amplitude at $A$ and $C$ sites is simply given by $a=-V_h c/V_v$ [see Fig.~\ref{f1}(b)-inset, where the scale goes from negative black to positive yellow, passing by a zero green amplitude]. This relation comes from the necessary amplitude balance in order to cancel the transport at the connector site $B$~\cite{luisFB}.

One interesting and important feature of FB states is that they can be excited in any region of the lattice, as soon as the zero amplitude condition at surrounding sites $B$ is fulfilled. The simultaneous excitation of compact states forms a coherent linear combination, which propagates stable along the $z$-direction. This allows us to transmit any combined pattern without any diffraction process, as the states forming this pattern have no diffraction at all (the slope of the FB is exactly zero). The simpler linear combination consists of two neighbor FB modes. They can be combined in phase or out of phase as Figs.~\ref{f1}(c1) and (c2) show, respectively. The main difference in terms of amplitudes is that the in-phase combination has a larger value at the central $A$-site, while the out of phase combination has a null amplitude at that position. In that sense, by linearly combining FB compact states we are able to compose different patterns that could be useful for codifying information as we will show below. 

\section{Experiments}
\begin{figure}[t]
\begin{center}
\includegraphics[width=0.47\textwidth]{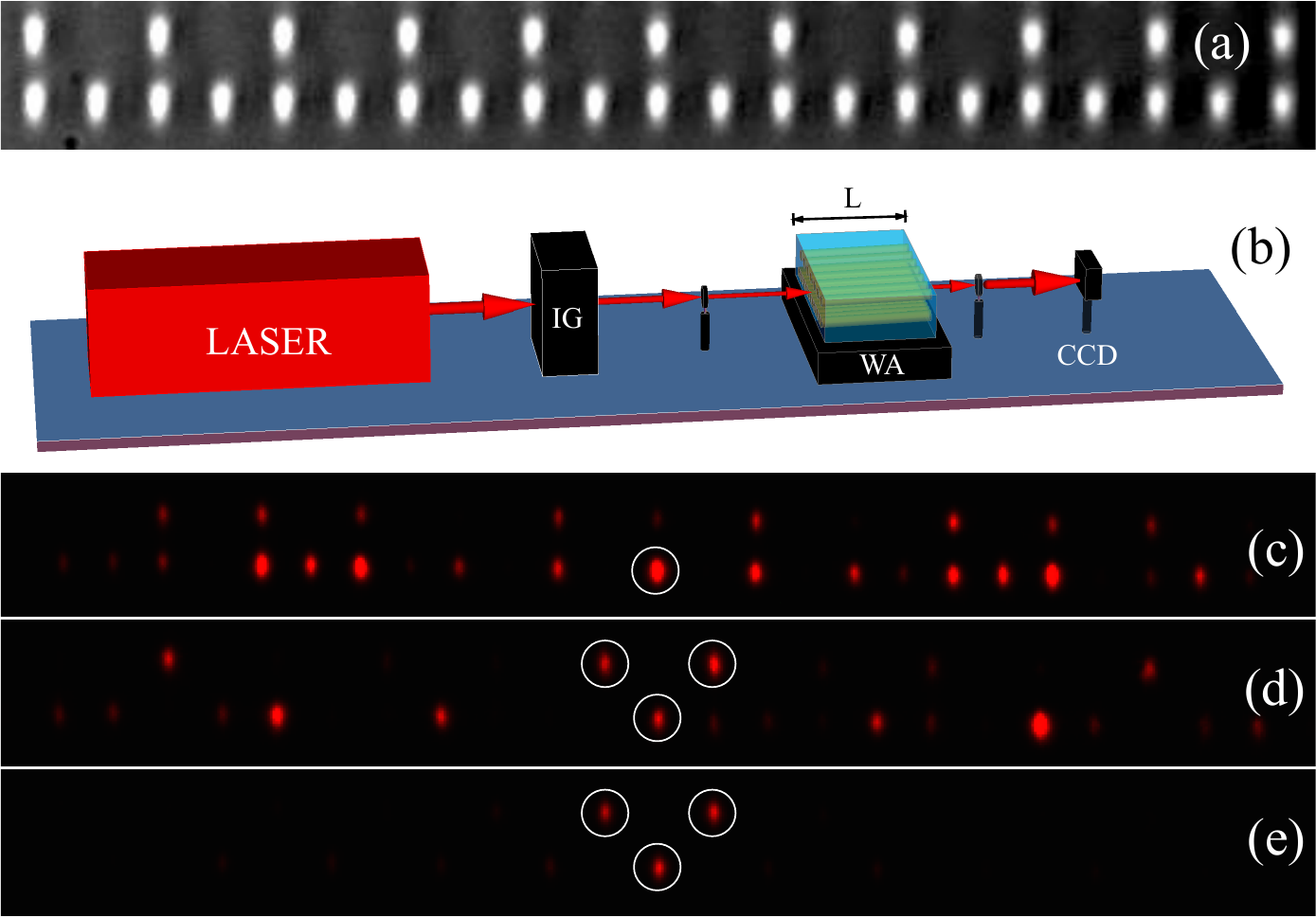}
\caption{(a) A microscope zoom of a Stub lattice. (b) Experimental setup. Output intensity profiles for different input excitations: (c) $B$-site, (d) three-sites in phase, and (e) three-sites out of phase. Circles in (c), (d) and (e) indicate the input positions.}\label{f2}
\end{center}
\end{figure}

In order to test the validity of model (\ref{model}) and particular features of the Stub geometry, we fabricated a Stub lattice using a femtosecond laser technique~\cite{fst}. Our lattice possesses a total of $77$ elliptical waveguides, with $51$ sites at the lower row and $26$ at the upper one. The waveguide spacing is $d_x=d_y=20\ \mu$m. To visually test the quality of our lattice, we launch white light at the input facet and take a picture of the output profile using a CCD camera, as shown in Fig.~\ref{f2}(a). We clearly see that although $d_x=d_y$, the waveguide ellipticity generates an effective anisotropy in terms of coupling constants~\cite{fst}; i.e, $V_v>V_h$. Fig.~\ref{f2}(b) shows a compact view of our experimental setup. We use an Image Generator (IG) configuration (composed of a sequence of Spatial Light Modulators, polarizers and optics) to transform a broad red laser beam of $633$ nm into a specific light pattern~\cite{lieb2}. Using this configuration, we are able to generate different initial conditions -- with amplitude and phase modulation -- which are then imaged at the lattice input facet. Then, light propagates though a $L=10$ cm long crystal and we obtain an image at the output facet by means of a standard CCD camera. 

We first study bulk transport in this lattice using a single-site excitation. This is performed by generating an image that illuminates an isolated bulk waveguide. In Fig.~\ref{f2}(c) we show the output spatial profile, after the excitation of a $B$-site at the bulk of the lattice (circle shows the input position). This initial condition only excites the dispersive part of the spectrum, due to the zero overlap with FB modes. We observe a rather symmetric transversal light distribution, including a characteristic discrete diffraction pattern~\cite{rep1}. Fig.~\ref{f2}(c) shows the larger spatial spreading observed exciting this lattice. The excitation of $A$ and $C$ sites shows a mixture between diffraction and localization, due to the excitation of the three linear bands.

A second experiment consists in exciting a Stub FB compact mode. First of all, we prepare an image that illuminates only three sites of the lattice, without any phase modulation [see Fig.~\ref{f2}(d)]. After a propagation of $10$ cm, we observe that some part of the energy remains at the input region but, also, some important part has diffracted across the lattice. This is expected because this input condition injects light at $A$ and $C$ sites and, therefore, a large part of the spectrum is effectively excited. Now, we use the same amplitude profile, but adding a staggered phase structure, mimicking the theoretical profile showed in Fig.~\ref{f1}(b)-inset. In Fig.~\ref{f2}(e) we show the excitation of this mode in a central region of the lattice. This image shows a dark background without any noticeable waveguide excited. This is a direct confirmation that the dynamics of this lattice is well described by model (\ref{model}) and that FB localized states exist stable on this lattice. Therefore, we are in good experimental conditions to propagate different composed Stub patterns and use this lattice as an optical code transmitter scheme~\cite{kagima,kagchen,lieb2,liebchen}.

\begin{figure}[t]
\begin{center}
\includegraphics[width=0.47\textwidth]{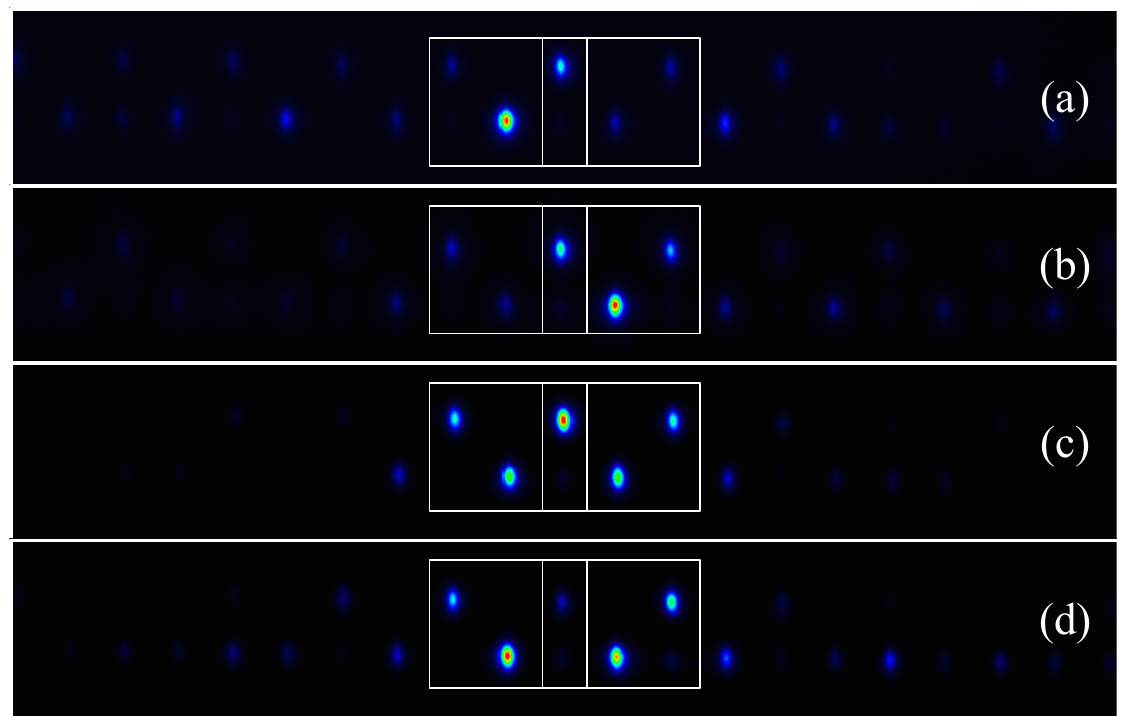}
\caption{Output intensity profiles for (a) $\phi_L$ and (b) $\phi_R$ states, and for (c) in phase and (d) out of phase FB mode combination.}\label{f3}
\end{center}
\end{figure}

Now, we combine two close FB states. As these modes have an amplitude and phase structure, we can combine them using in phase and out of phase composition. The FB state located to the left is called $\phi_L$ [see Fig.~\ref{f3}(a)] and the one located to the right is called $\phi_R$ [see Fig.~\ref{f3}(b)]. We could generate infinite linear combinations depending on the coefficients in front of these states: $\alpha\phi_L+\beta \phi_R$, with $\alpha,\beta\in \mathbb{R}$. This linear combination will be coherent along the propagation coordinate $z$, because both states belong to the same band, having the same propagation constant (frequency) $\beta=0$. For simplicity, we consider two symmetric linear combinations: $\phi_L\pm \phi_R$ [as sketched in Fig.~\ref{f1}(c)]. Experimentally, we study these composed states by generating input patterns having the corresponding amplitude and phase modulation. Figs.~\ref{f3}(c) and (d) show the output profiles for in phase and out of phase combinations, respectively, after a propagation of $10$ cm. Although we observe very low intensity peaks at surrounding sites (due to the high contrast of these images), this is not affecting the predominant composed profile (the input image composition could always have some asymmetries in amplitudes and phases, that could excite part of the dispersive spectrum). When there is an in phase combination, we observe the constructive interference at the central upper site as showed in Fig.~\ref{f3}(c), with a very large intensity at this position. On the other hand, we observe a destructive interference at this site for an out of phase linear combination, as showed in Fig.~\ref{f3}(d). 

\begin{figure}[h]
\begin{center}
\includegraphics[width=0.47\textwidth]{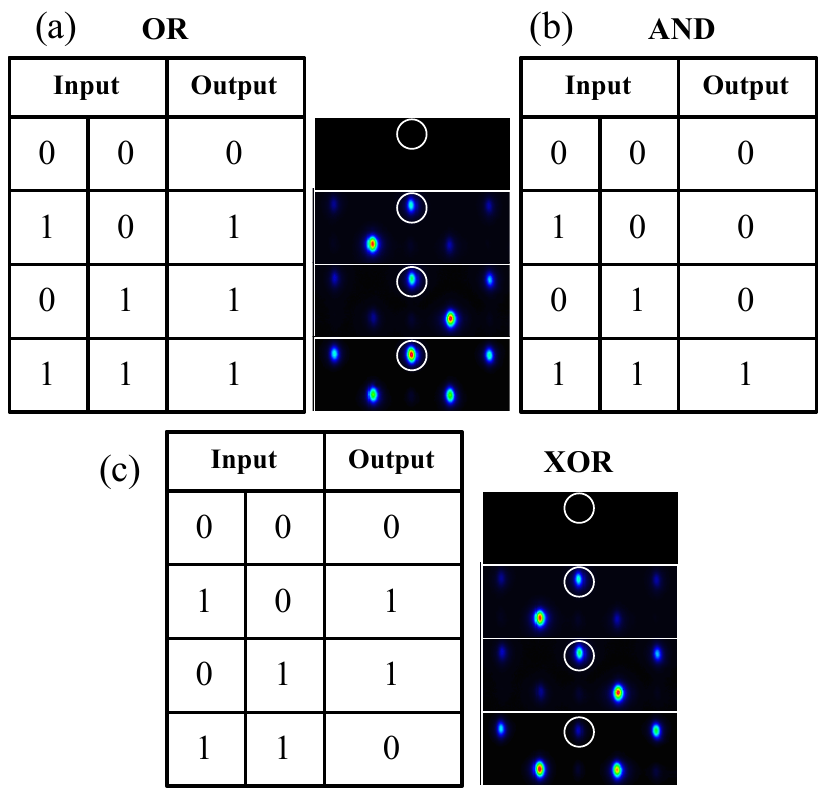}
\caption{Truth tables for (a) OR, (b) AND and (c) XOR gates. Output experimental images are also showed, together with their corresponding inputs from the tables.}\label{f4}
\end{center}
\end{figure}

In Fig.~\ref{f3} we observe three intensity levels at the central upper position site: low (background level), middle (single excitation) and large (in-phase combination) intensities. These three options give us the possibility to use the combination of $\phi_L$ and $\phi_R$ states to perform a number of well known logical operations. We use these states as left and right input signals in our gates, and define the intensity at the center top site as the output channel. For example, an ``OR gate'' gives $0$ only when no signal is measured at the output. Therefore, as showed in Fig.~\ref{f4}(a), exciting only $\phi_L$ or only $\phi_R$ or an in phase combination $\phi_L+ \phi_R$, we exactly generate an OR gate using the FB modes of our Stub lattice. 

As the intensity at the output channel (indicated by a circle) could have a low, middle or large value, we can also differentiate at the output by measuring this amount. To generate an ``AND gate'', we use exactly the same inputs as before, but now defining a threshold for the output to be considered as $1$. We simply define that middle or lower intensities correspond to a $0$, while the large intensity generated by $\phi_L+ \phi_R$ is defined as an output $1$ [see Fig.~\ref{f4}(b)].

Finally, we generate a ``XOR gate'' by using the same inputs than before, but now considering an out of phase linear combination $\phi_L- \phi_R$ [see Fig.~\ref{f4}(c)]. Therefore, when measuring a middle intensity at the center top site, we will define an output equal to $1$. This will occur when inputs are injected only independently, as XOR gate works. If no modes or both are injected in an out of phase linear combination, the center top site will have a zero or very low amplitude, corresponding to a $0$ at the output channel.

We could increase the performance of our system by re-defining the output channel measurement. Although it is simpler to define an output ``0'' when there is no light and an output ``1'' when detecting some amount of it, we could just define our system scheme in the opposite way, just defining that no light means an output ``1'' and light (depending on its level for an AND gate) corresponds to a ``0''. This could be implemented just by definition in our detection scheme at the output channel position (camera, photodiode, etc.), or by adding a negation gate after the already described system. By doing this, we would be in conditions of generating NOR, NAND and NXOR gates as well.

\section{Conclusions}

In conclusion, we have been able to experimentally excite the linear FB modes of a Stub photonic lattice. We have clearly contrasted the localization features of this state with respect to bulk transport. The linear superposition of two neighbor FB modes was observed, considering an in phase and out of phase combination.  We proposed a novel basic design of all-optical logic gates by demonstrating the use of FB Stub states as input channel to perform three logical operations, based on a single and combined excitation. Our design is still in an early phase of the development and evaluation of its numerical properties, such as operating speed and integration capacity. Our experiments were performed on a $L=10$ cm long crystal having $77$ waveguides, but similar results may be observable in shorter and smaller arrays (the FB localization properties rely on very discrete features of the lattice, where only few unitary cells are necessary for observing the described phenomenology~\cite{luisFB}). This is a very important aspect when considering a reduction of the system size for concrete applications. Additionally, the femtosecond technique~\cite{fst} also allows the possibility to write complex arrays in order to perform concatenated operations~\cite{aipalex}. Our proposed all-optical logic gates, therefore, have promising potentials to produce core units to implement various all-optical systems for optical signal processing.

\subsection{Acknowledgments}

This work was supported in part by Programa ICM grant RC130001 and FONDECYT Grant No. 1151444.



\end{document}